\documentstyle[12pt,fleqn,epsf]{article}

\newcommand{\ra}{\rightarrow}
\newcommand{\eq}{\begin{equation}}
\newcommand{\en}{\end{equation}}
\newcommand{\eqa}{\begin{eqnarray*}}
\newcommand{\ena}{\end{eqnarray*}}

\newcommand{\bt}{\beta}

\newcommand{\caM}{{\cal M}}
\newcommand{\caD}{{\cal D}}
\newcommand{\bpsi}{{\bar \psi}}

\newcommand{\AmS}{{\protect\the\textfont2
  A\kern-.1667em\lower.5ex\hbox{M}\kern-.125emS}}

\begin{document}
\hbox{}\hfill {\small HUB-EP-97/70}\\

\begin{center}
\vspace*{1.0cm}

{\LARGE
Effects of dynamical Wilson fermions and the phase structure of
compact QED$_{\rm 4}\mbox{}$\footnote{Contribution to LATTICE 97 --- XVth Int. 
Symp. on Lattice Field Theory, Edinburgh, Scotland.}}

\vspace*{0.5cm}
{\large
A. Hoferichter$\mbox{}^a$,
V.K. Mitrjushkin$\mbox{}^b$,
M. M\"uller-Preussker$\mbox{}^c$
and H. St\"uben$\mbox{}^d$
}
                          
\vspace*{0.2cm}
{\normalsize
{$\mbox{}^a$ \em DESY-IfH and HLRZ, Zeuthen, Germany}\\
{$\mbox{}^b$ \em Joint Institute for Nuclear Research, Dubna, Russia}\\
{$\mbox{}^c$ \em Institut f\"ur Physik,Humboldt-Universit\"at zu Berlin, Germany}\\
{$\mbox{}^d$ \em Konrad-Zuse-Zentrum f\"ur Informationstechnik Berlin, Germany}
}

\vspace{1cm}
{\bf Abstract}
\end{center}
By comparison of the quenched and full formulations
of compact QED with Wilson fermions we single out
the effects of dynamical fermions on the `chiral
transition line' within the confinement phase. It is shown that this line
cannot correspond to the chiral limit of the theory
for all values of the gauge coupling.
This seems to imply the existence of tri-critical points
in this theory, the phase structure of which has a close
similarity to QCD at finite temperature.

\section{Introduction}

Wilson lattice fermions break chiral symmetry explicitly.
Hopefully, it can be recovered by fine-tuning the bare parameters
in the continuum limit. In the phase diagram the `critical' line 
$~\kappa_c(\bt)~$  is associated with the chiral limit of the theory.
For non-vanishing lattice spacing on this line only a partial
restoration of chiral symmetry occurs.

In this letter we are concerned with the behavior of 
fermionic observables, in particular with the pseudo-scalar 
`pion' mass close to $~\kappa_c(\bt)~$ in the confinement 
phase of compact QED. We confront the
full theory with its valence approximation.

\section{The Model, Observables and the Phase Structure}
We consider the Wilson action for QED
\eq
 S_{W} = S_{G}(U) + S_{F}(U, {\bar \psi}, \psi)
\en
consisting of the standard plaquette compact $U(1)$
gauge action $S_{G}(U)$
and the fermionic part
\eq
S_{F}(U, {\bar \psi}, \psi) =  \sum_{f=1,2}\sum_{x,y}
\bpsi_x^f \caM_{xy}(U) \psi_y^f~,
\en
with the Wilson matrix $\caM (U)= {\hat 1}-\kappa \caD (U),$
\eq
\caD_{xy} \equiv  \sum_{\mu} \Bigl[ \delta_{y, x+\hat{\mu}}
P^{-}_{\mu} U_{x \mu}
+ \delta_{y, x-\hat{\mu}} P^{+}_{\mu} U_{x-\hat{\mu},\mu}^{\dagger} \Bigr]
\en
and $P^{\pm}_{\mu}={\hat 1}\pm\gamma_{\mu}$.
We adopt the relation of the hopping parameter $\kappa$ to the fermion mass
$~m_q=(1/\kappa - 1/\kappa_c(\bt))/2$. 

\begin{figure}[htb]
\vspace{-0.5cm}

\epsfysize=14.2cm
\epsfxsize=14.2cm
\epsfbox{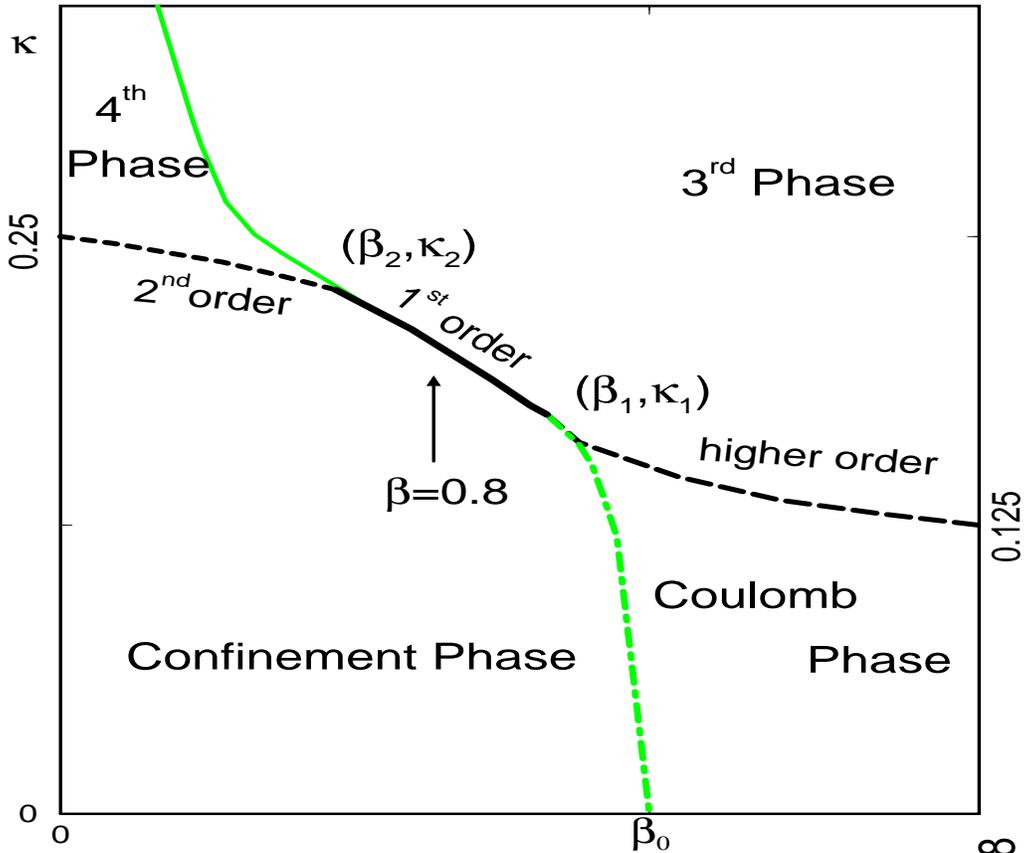}

\vspace{-2.0cm}
\caption{The phase diagram of compact lattice QED with dynamical
         Wilson fermions}
\label{fig:0col}
\end{figure}

The phase structure of the path integral quantized theory has been
investigated in \cite{qed2}. For full QED a diagram with four phases in the
range $~0 \le \kappa < 0.3~$ emerged (see Fig. \ref{fig:0col}).
Whereas the Coulomb and confinement phases are quite well
understood, the situation in the `upper' areas deserves further
study (see e.g. \cite{aoki}). We mention
that this phase diagram  has similarities with that of lattice 2-flavor 
QCD at finite $~T~$ 
\cite{ukawa}. 

Our previous investigations were mainly restricted  
to fermionic bulk variables like
$<{\overline \psi}\psi>$, $<{\overline \psi}\gamma_5\psi>$ and
the `pion' norm 
$$\langle \Pi \rangle
 =  \Bigl\langle \mbox{Tr} \Bigl( {\cal M}^{-1}
\gamma_{5} {\cal M}^{-1} \gamma_{5} \Bigr) \Bigr\rangle_{G}/4V.$$
$~\langle ~~ \rangle_{G}~$ means averaging over gauge
field configurations,  $~V=N_{\tau} \cdot N_s^3~$  is the number of 
lattice sites.

Here we present mainly results for the `pion' mass $m_{\pi}$ obtained from
the pseudo-scalar non-singlet correlator
\eq
\Gamma (\tau )
 \equiv  \frac{1}{N_s^6} \cdot \sum_{\vec{x},\vec{y}}~
\left\langle \mbox{Sp}
\Bigl( {\cal M}^{-1}_{x y} \gamma_5 {\cal M}^{-1}_{y x} \gamma_5 \Bigr)
\right\rangle_{G}~,
                              \label{pscorr}
\en 
where ~~Sp~~ denotes the trace with respect to the Dirac indices.

The simulations for full QED  have been carried out with the
HMC method, the inversion of the matrix $~{\cal M}~$
with a BiCGstab algorithm.
As in our previous work \cite{qed2} for
representative $~\bt$--values within the confinement phase we have 
considered the strong coupling limit $~\bt=0$ and 
$~\bt=0.6,~0.8$. 
\section{Results}
First let us discuss the strong coupling limit $\beta=0$. There, for both
the quenched approximation and the full theory
the `pion norm' data are compatible with a PCAC-like relation
\eq
\langle \Pi \rangle = \frac{C_0}{m_q} + C_1
~,\quad m_q \ra 0~,
                                            \label{Pole_mpi}
\en
where the constant $~C_0 > 0~$ -- up to a factor -- can be identified with
the subtracted chiral condensate (see \cite{hmm95}).
$C_0~$ and $~C_1~$ turned out to differ slightly for the quenched and
dynamical cases. The fits for $~\kappa_c(0)~$ provided for the quenched case 
$~\kappa_c(0) = 0.2502(1)~$ (nicely agreeing with the analytically known
value $~1/4~$) and for the dynamical case a slightly shifted value 0.2450(6). 

\begin{figure}[htb]
\epsfysize=12.0cm
\epsfxsize=12.0cm
\epsfbox{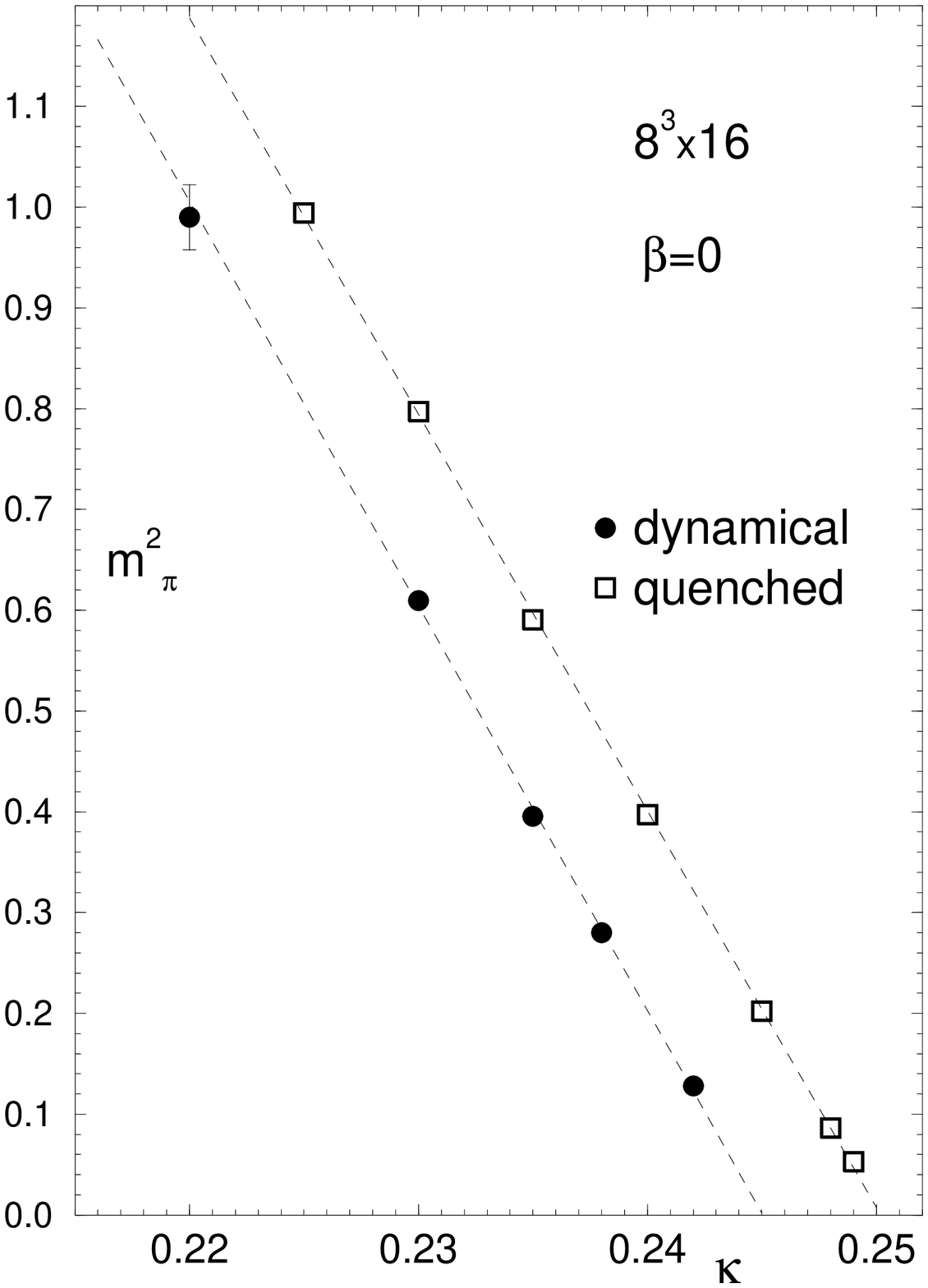}
\vspace{-0.5cm}
\caption{$m_{\pi}^2$ as function of $\kappa$ for the quenched and full 
compact QED at $\beta=0$. }
\label{fig:4col}
\end{figure}

In Fig. \ref{fig:4col} the dependence of
$~m_{\pi}^2~$ on $~\kappa~$ for the full and
quenched theories is shown for an  $~8^3\times 16~$ lattice
at $~\bt=0$. The quenched data for $~\kappa$ very close to $~\kappa_c~$
were obtained by an improved estimator
of $~m_{\pi}~$ \cite{hmm95} in order to increase the
signal-to-noise ratio.
%
%
%
In both, quenched and dynamical cases we observe a
dependence of $~m^2_{\pi}~$ on $~\kappa~$ compatible with
$m^2_{\pi} \sim \Bigl( 1 - \frac{\kappa}{\kappa_c}\Bigr) ~,\quad
\kappa \leq \kappa_c~,$
which in the limit $~\kappa \rightarrow \kappa_c~$ transforms into the
PCAC--like relation 
$ m_{\pi}^2 = B \cdot m_q ~,\quad m_q \ra 0+$.
The linear extrapolations of $~m_{\pi}^2~$ down to zero provide
estimates for $~\kappa_c(0)~$ compatible with the above mentioned values.
Thus, we conclude that at very strong coupling a chiral
limit with a vanishing pseudo-scalar mass can be defined both in the quenched
and the dynamical fermion case. The chiral transition at $~\beta = 0~$
is of second order.

\begin{figure}[htb]
\vspace{-9pt}
\hspace*{0.8cm}
\epsfysize=12.4cm
\epsfxsize=11.8cm
\epsfbox{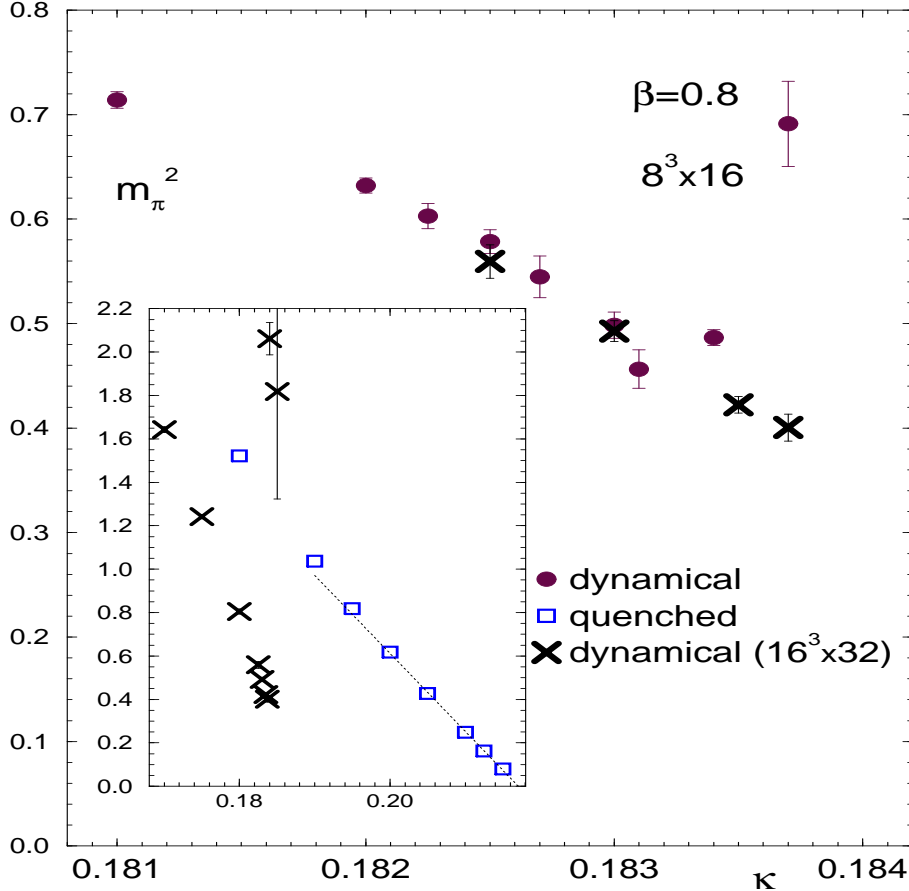}
\vspace{-0.4cm}
\caption{Same as Fig.\protect\ref{fig:4col} but at $\beta=0.8$. Note
the scales on the inset and outer plot.}
\label{fig:5col}
\end{figure}

The situation changes drastically, if we proceed to larger  $~\beta$--values.
At $~\beta=0.8~$ the quenched and dynamical cases strongly differ from
each other. In the quenched case the behaviour of the
`pion norm' and of the `pion' mass very much resembles to the results obtained
in the strong coupling limit $~\beta=0~$. 
However, in the presence of dynamical fermions the fermionic bulk variables 
as well as gauge observables undergo a discontinuous jump
at $~\kappa_c(0.8) = 0.1832(3)~$.
The discontinuity of the `pion norm' increases with the lattice size
as has been checked for sizes $~8^3\times 16~$ and $~16^3\times 32~$.
On top of $~\kappa_c~$ a clear metastability behaviour was observed. 
The `pion' mass passes a non-vanishing minimum value at $~\kappa_c~$.
It only slightly decreases for increasing lattice sizes. The
corresponding data are plotted in
Fig. \ref{fig:5col}. 

%
%
%

We conclude that at $~\beta~= 0.8~$ 
the transition becomes certainly a first order transition under the
influence of the fermionic determinant. As far as the pseudo-scalar
mass does not vanish the standard definition of the chiral limit 
does not apply.

We carried out simulations also at $~\beta~= 0.6~$. 
There the transition becomes weaker, the minimal `pion' mass decreases. 
This is not surprising, because we are approaching a second order transition 
at stronger bare coupling. 

\section{Summary}
We have studied the approach to $~\kappa_c(\beta)~$ for several 
$~\bt$--values within the confinement phase of the 
compact lattice QED with Wilson fermions comparing the
full theory with its quenched approximation.
We have shown the importance of vacuum polarization effects
due to dynamical fermions in the context of the chiral limit.
In the strong coupling limit $~\bt=0$ the
only effect of dynamical fermions seems to be a renormalization
of the `critical' value $~\kappa_c$,~$~\kappa_c^{dyn}
   \not = \kappa_c^{quen}$.
In the quenched as well as dynamical fermion case
the pseudo-scalar particle
becomes massless when $~\kappa \ra \kappa_c$.
However, at $~\beta=0.8~$ the presence of the dynamical (`sea') fermions
drastically changes the transition. There
the transition cannot be anymore associated with the zero-mass limit of
a pseudo-scalar particle, in sharp contrast to the quenched case.
Since the transition is first order, we can speculate
about the existence of tri-critical points on the line $\kappa_c(\beta)$.

\end{document}